\documentclass[a4paper]{article}

\usepackage[utf8]{inputenc}
\usepackage[top=10mm]{geometry}
\usepackage{booktabs}
\usepackage[flushleft]{threeparttable}
\usepackage{authblk}
\usepackage{bm}
\usepackage{amsmath}
\usepackage{amssymb}
\usepackage{graphicx}
\usepackage[square,sort,numbers]{natbib}
\allowdisplaybreaks[4]

\title{Pion-nucleon scattering to order $p^4$ in SU(3) heavy baryon chiral perturbation theory}

\author[1]{Bo-Lin Huang \thanks{bolin.huang@foxmail.com}}

\affil[1]{\normalsize Department of Physics, Jishou University, Jishou 416000, China}

\date{\today}

\begin{document}
\maketitle

\begin{abstract}
  We calculate the $T$-matrices of elastic pion-nucleon ($\pi N$) scattering up to fourth order in
  SU(3) heavy baryon chiral perturbation theory. The pertinent low-energy constants are determined by fitting to $\pi N$ phase shifts below 200 MeV pion laboratory momentum in the physical region. The scattering lengths and scattering volumes are extracted from the chiral amplitudes, and  turn out to be in good agreement with those of other approaches and the available experimental values. We also discuss the subthreshold parameters and the related issues. On the basis of the various phase shifts, the threshold parameters and the subthreshold parameters, the convergence of the chiral expansion is analyzed in detail. The calculation provides the possibility to consider explicitly more complex processes involving strangeness.
\begin{description}
\item[PACS numbers:]
13.75.Jz,12.39.Fe,12.38.Bx
\item[Keywords:]
Chiral perturbation theory, pion-nucleon scattering, chiral convergence
\end{description}
\end{abstract}

\section{Introduction}
Chiral perturbation theory (ChPT) allows one to analyze hadronic processes at low energies,  that are not accessible by a perturbative expansion in the strong coupling constant $\alpha_s$ of quantum chromodynamics (QCD)\cite{wein1979,gass1984,leut1994,sche2012}. ChPT is an efficient framework to calculate in a model-independent way, e.g., the amplitudes of pion-nucleon ($\pi N$) scattering below the chiral symmetry breaking scale $\Lambda_{\chi}\simeq 1 \text{GeV}$. However, there exists a power-counting problem in baryon ChPT because of the non-vanishing baryon mass $M_0$ in the chiral limit. Over the years, several approaches have been proposed to solve this  problem. Heavy baryon ChPT (HB$\chi$PT) \cite{jenk1991,bern1992}, the infrared regularization of covariant baryon ChPT \cite{bech1999}, and the extended-on-mass-shell
scheme for baryon ChPT \cite{gege1999,fuch2003} are some popular approaches. The last two approaches are fully relativistic and have lead to substantial progress in many aspects as documented in ref.~\cite{schi2007,geng2008,mart2010,ren2012,alar2012,alar2013,geng2013}. However, the expressions from the loop diagrams become rather complicated in these fully relativistic approaches \cite{chen2013,yao2016,lu2019}. On the other hand, HB$\chi$PT is a well-established and versatile tool for the study of low-energy hadronic processes. The amplitudes in HB$\chi$PT proceeds simultaneously in terms of $p/\Lambda_\chi$ (non-relativistic contributions from tree- and loop-diagrams) and $p/M_0$ (relativistic corrections), where $p$ denotes the meson momentum (or mass) or the small residual momentum of a baryon in a low-energy process.

In recent years, there has been renewed interest in theoretical studies of elastic meson-baryon scattering at low energies. These are not only concerned with the description of the strong mesonic interaction, but also with the chiral properties of the baryons. The low-energy processes between pions and nucleons have been investigated extensively in  SU(2) HB$\chi$PT in refs.~\cite{fett1998,fett2000,kreb2012,ente2015}. Furthermore, the subthreshold parameters of pion-nucleon scattering have also been deeply studied by combining the Roy-Steiner (RS) equations and SU(2) ChPT \cite{hofe20151,siem2017}. For processes involving kaons or hyperons, the situation becomes more involved, since one has to work out the consequences of three-flavor chiral dynamics. In a previous paper \cite{huan2015} we have investigated $KN$ and $\bar{K}N$ elastic scattering up to one-loop order in SU(3) HB$\chi$PT by fitting low-energy constants to partial-wave phase shifts of $KN$-scattering and obtained quite reasonable results. This approach was then extended by predicting the amplitudes of pseudoscalar-meson octet-baryon scattering in all channels, with the pertinent low-energy constants fitted to partial-wave phase shifts of elastic $\pi N$ and $KN$ scattering \cite{huan2017}. At this point one should note that the detailed predictions  of SU(3) HB$\chi$PT for the meson-baryon scattering lengths have been given previously in refs.~\cite{kais2001,liu20071,liu20072,liu2011,liu2012}. Moreover, these studies  in SU(3) HB$\chi$PT has been extended to partial-wave phase shifts, the pion-nucleon sigma term, and other quantities.  In a recent paper \cite{huan2020}, we have calculated the complete $T$-matrices of pion-nucleon scattering up to third order in SU(3) HB$\chi$PT and obtained as a byproduct nucleon properties, like the pion-nucleon sigma term $\sigma_{\pi N}$. A good description of the phase shifts of $\pi N$ scattering below 200 MeV has been obtained, and then it can serve as a consistency check to consider also the other meson-baryon scattering channels. However, no good convergence of the chiral expansion is observed in calculations that terminate at third order. Therefore, we will compute in this paper the $T$-matrices for $\pi N$ scattering up to fourth order in SU(3) HB$\chi$PT. The pertinent low-energy constants (LECs) will be determined by fitting to empirical $S$- and $P$-wave phase shifts of $\pi N$ scattering. In particular, the LECs related to contact terms of chiral dimension will be obtained separately. The threshold parameters, the subthreshold parameters and the related issues will be discussed briefly. At last, the convergence of the chiral expansions will also be analyzed and discussed in detail.

The present paper is organized as follows. In Sec.~\ref{lagrangian}, we summarize the Lagrangians involved in the evaluation of the fourth-order contributions. In Sec.~\ref{tmatrices}, we present some explicit expressions for the $T$-matrices of elastic $\pi N$ scattering at order $p^4$. In Sec.~\ref{phase}, we outline how to derive phase shifts, scattering lengths, and scattering volumes from the $T$-matrices. Section~\ref{results} contains the presentation and discussion of our results and it also includes a brief summary. Appendixes~\ref{scattering lengths} and~\ref{subthresholdexpressions} contain the expressions for the scattering lengths and the subthreshold parameters, respectively.

\section{Chiral Lagrangian}
\label{lagrangian}
In order to calculate the pion-nucleon scattering amplitude up to order $p^4$ in SU(3) heavy baryon  chiral perturbation theory, one has to evaluate tree and loop diagrams with vertices from the effective chiral Lagrangian:
\begin{align}
\label{eq1}
\mathcal{L}_{\text{eff}}=\mathcal{L}^{(2)}_{\phi\phi}+\mathcal{L}^{(1)}_{\phi B}+\mathcal{L}^{(2)}_{\phi B}+\mathcal{L}^{(3)}_{\phi B}+\mathcal{L}^{(4)}_{\phi B}.
\end{align}
These Lagrangians are written in terms of the traceless Hermitian $3\times 3$ matrices $\phi$ and $B$ that include the pseudoscalar Goldstone boson fields ($\pi$, $K$, $\bar{K}$, $\eta$) and
the octet-baryon fields ($N$, $\Lambda$, $\Sigma$, $\Xi$), respectively. The explicit form of the $\mathcal{L}^{(2)}_{\phi\phi}$, $\mathcal{L}^{(1)}_{\phi B}$, $\mathcal{L}^{(2)}_{\phi B}$, and $\mathcal{L}^{(3)}_{\phi B}$ can be found in ref.~\cite{huan2020}. The complete fourth-order heavy baryon Lagrangian $\mathcal{L}^{(4)}_{\phi B}$ naturally splits up into two parts: relativistic corrections with fixed coefficients, and counterterms proportional to new low-energy constants. The relativistic terms can be obtained from the original leading order, next-to-leading order and next-to-next-to-leading order lorentz-invariant Lagrangians through path integral manipulations \cite{bern1992}. For three-flavors and at chiral order four, the Lorentz-invariant meson-baryon Lagrangian has been constructed in ref.~\cite{jian2017} and we can obtain from it the counterterms relevant for our purpose. Since the dimension-four chiral meson-baryon Lagrangian  $\mathcal{L}^{(4)}_{\phi B}$ is very lengthy, the explicit expressions will not be given in this paper, where we consider only elastic pion-nucleon scattering. Of course, the expansions in SU(3) and SU(2) HB$\chi$PT are consistent with each other. For this reason the notation of low-energy constants $\bar{e}_i(i=14,...,22,35,...,38)$ for the fourth-order counterterms in SU(2) HB$\chi$PT introduced in ref.~\cite{fett2000} will also be used in this paper.

\section{$T$-matrices for pion-nucleon scattering}
\label{tmatrices}
We are considering the elastic pion-nucleon scattering process $\pi(q)+N(p) \rightarrow \pi(q')+N(p')$ in the center-of-mass system. In states with total isospin  $I=1/2$ or $I=3/2$ of the pion-nucleon system, the corresponding $T$-matrix takes the following form in spin space:
\begin{align}
\label{eq12}
 T_{\pi N}^{(I)}=V_{\pi N}^{(I)}(w,t)+i\bm{\sigma}\cdot(\bm{q}'\times\bm{q})\,W_{\pi N}^{(I)}(w,t)
\end{align}
where $w=q_0=q'_0=(m_\pi^2+\bm{q}^2)^{1/2}$ is the pion center-of-mass energy, and $t=(q'-q)^2=2\bm{q}^2(z-1)$ is the invariant momentum transfer squared with $z=\cos\theta$ the cosine of the angle $\theta$ between $\bm{q}$ and $\bm{q}'$. Based
on relativistic kinematics, one gets the relation:
\begin{align}
\label{eq13}
&\bm{q}'^{2}=\bm{q}^2=\frac{M_{N}^{2}\bm{p}_{\text{lab}}^{2}}{m_{\pi}^{2}+M_{N}^{2}+2M_{N}\sqrt{m_{\pi}^{2}+\bm{p}_{\text{lab}}^{2}}},
\end{align}
where $\bm{p}_{\text{lab}}$ denotes the momentum of the incident meson in the laboratory system. Furthermore, $V_{\pi N}^{(I)}(w,t)$ refers to the non-spin-flip pion-nucleon scattering amplitude and $W_{\pi N}^{(I)}(w,t)$ is called the spin-flip pion-nucleon scattering amplitude.

\begin{figure}[t]
\centering
\includegraphics[height=12cm,width=8cm]{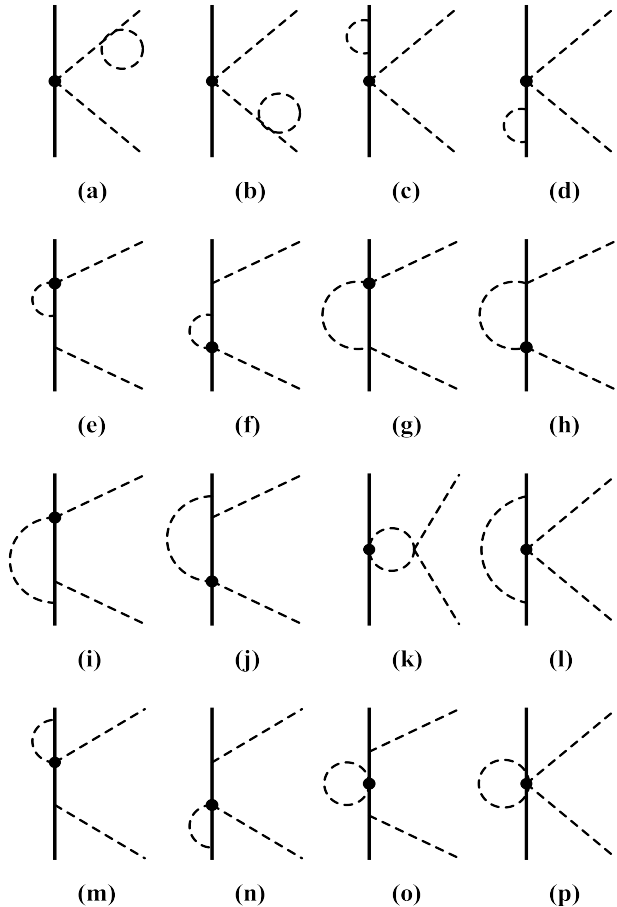}
\caption{\label{fig:oneloopfeynman}Nonvanishing one-loop diagrams contributing at chiral order four. The heavy dots refer to vertices from $\mathcal{L}_{\phi B}^{(2)}$. Crossed graphs are not shown.}
\end{figure}

From leading order $\mathcal{O}(p)$ up to third order $\mathcal{O}(p^3)$, the expressions for the amplitudes  $V_{\pi N}^{(I)}(w,t)$ and $W_{\pi N}^{(I)}(w,t)$ can be found in ref.~\cite{huan2020}. At fourth order $\mathcal{O}(p^4)$,  the SU(3) results for the $\pi N$ amplitudes from tree diagrams are essentially the same as those calculated in SU(2) HB$\chi$PT in ref.~\cite{fett2000}. After an appropriate renaming of the low-energy constants, these contributions read
\begin{align}
\label{eq14}
V_{\pi N}^{(3/2,\text{N3LO})}=&-\frac{(D+F)^2}{32M_0^3w^4f_\pi^2}[(t^4+7t^3w^2+11t^2w^4-3tw^6+4w^8)-(11t^3+49t^2w^2+32tw^4\nonumber\\
&+4w^6)m_\pi^2+(45t^2+110tw^2+26w^4)m_\pi^4-3(27t+26w^2)m_\pi^6+54m_\pi^8]\nonumber\\
&+\frac{m_\pi^2-w^2}{32M_0^3f_\pi^2}(4m_\pi^2-t-4w^2)+\frac{1}{16M_0^2f_\pi^2}[4C_0(t+2w^2-2m_\pi^2)m_\pi^2+C_1(2t^2+4tw^2\nonumber\\
&-8tm_\pi^2-8w^2m_\pi^2+8m_\pi^4)+4C_2(3tw^2+14w^4-4tm_\pi^2-22w^2m_\pi^2+8m_\pi^4)\nonumber\\
&+2C_3(-t^2-4tw^2+4tm_\pi^2)]-\frac{8}{M_0 f_\pi^2}C_0C_2w^2m_\pi^2+\frac{1}{2M_0 f_\pi^2}[H_1(4m_\pi^2-t-4w^2)m_\pi^2\nonumber\\
&+H_2(4w^2+t-4m_\pi^2)t+3H_3(4m_\pi^2-t-4w^2)w^2+2H_4w^2t]\nonumber\\
&+\frac{1}{f_\pi^2}[4\bar{e}_{14}(4m_\pi^4-4m_\pi^2 t-t^2)+8\bar{e}_{15}(2w^2m_\pi^2-tw^2)+16\bar{e}_{16}w^4],
\end{align}
\begin{align}
\label{eq15}
W_{\pi N}^{(3/2,\text{N3LO})}=&\frac{(D+F)^2}{16M_0^3w^4f_\pi^2}[(t^3+5t^2w^2+3tw^4-w^6)-(9t^2+25tw^2+4w^4)m_\pi^2\nonumber\\
&+3(9t+10w^2)m_\pi^4-27m_\pi^6]+\frac{w^2-m_\pi^2}{16M_0^3 f_\pi^2}+\frac{1}{8M_0^2f_\pi^2}[8C_0m_\pi^2+C_1(2t-4m_\pi^2)\nonumber\\
&-4C_2w^2-2C_3(t+2w^2-2m_\pi^2)]+\frac{1}{2M_0 f_\pi^2}H_4(8w^2+t-4m_\pi^2)\nonumber\\
&+\frac{1}{f_\pi^2}[\bar{e}_{17}(-8m_\pi^2+4t)-8\bar{e}_{18}w^2],
\end{align}
\begin{align}
\label{eq16}
V_{\pi N}^{(1/2,\text{N3LO})}=&\frac{(D+F)^2}{64M_0^3w^4f_\pi^2}[(t^4+7t^3w^2+11t^2w^4+16w^8)-(11t^3+49t^2w^2+32tw^4\nonumber\\
&+16w^6)m_\pi^2+5(9t^2+22tw^2+4w^4)m_\pi^4-84(t+w^2)m_\pi^6+60m_\pi^8]\nonumber\\
&+\frac{w^2-m_\pi^2}{16M_0^3f_\pi^2}(4m_\pi^2-t-4w^2)+\frac{1}{8M_0^2f_\pi^2}[2C_0(t+2w^2-2m_\pi^2)m_\pi^2+C_1(t^2+2tw^2\nonumber\\
&-4tm_\pi^2-4w^2m_\pi^2+4m_\pi^4)+2C_2(3tw^2+14w^4-4tm_\pi^2-22w^2m_\pi^2+8m_\pi^4)\nonumber\\
&+2C_3(t^2+4tw^2-4tm_\pi^2)]-\frac{8}{M_0 f_\pi^2}C_0C_2w^2m_\pi^2+\frac{1}{M_0 f_\pi^2}[H_1(-4m_\pi^2+t+4w^2)m_\pi^2\nonumber\\
&-H_2(4w^2+t-4m_\pi^2)t-3H_3(4m_\pi^2-t-4w^2)w^2+H_4w^2t]\nonumber\\
&+\frac{1}{f_\pi^2}[4\bar{e}_{14}(4m_\pi^4-4m_\pi^2 t-t^2)+8\bar{e}_{15}(2w^2m_\pi^2-tw^2)+16\bar{e}_{16}w^4],
\end{align}
\begin{align}
\label{eq17}
W_{\pi N}^{(1/2,\text{N3LO})}=&\frac{(D+F)^2}{32M_0^3w^4f_\pi^2}[(-t^3-5t^2w^2-3tw^4+4w^6)+(9t^2+25tw^2+4w^4)m_\pi^2\nonumber\\
&-3(9t+10w^2)m_\pi^4+24m_\pi^6]+\frac{m_\pi^2-w^2}{8M_0^3 f_\pi^2}+\frac{1}{4M_0^2f_\pi^2}[-2C_0m_\pi^2+C_1(t-2m_\pi^2)-2C_2w^2\nonumber\\
&+2C_3(t+2w^2-2m_\pi^2)]+\frac{H_4}{2M_0 f_\pi^2}(8w^2+t-4m_\pi^2)+\frac{8}{f_\pi^2}[\bar{e}_{17}(2m_\pi^2-t)+2\bar{e}_{18}w^2],
\end{align}
where $C_i\, (i=0,1,2,3)$ and $H_i\, (i=1,2,3,4)$ are linear combinations of low-energy constants definded in eq.(22) and eq.(27) of ref.~\cite{huan2020}, respectively. For the dimension-four LECs, we follow the same strategy as in ref.~\cite{fett2000}. The subset of dimension-four low-energy constants $\bar{e}_i\,(i=19,20,21,22,35,36,37,38)$ can be absorbed onto the dimension-two LECs $c_{i}\,(i=1,2,3,4)$, see eq.(3.23) in ref.~\cite{fett2000}. We note that, the dimension-two LECs $c_i\,(i=1,2,3,4)$ introduced in SU(2) HB$\chi$PT can be replaced equivalently by the combinations of LECs $C_i\, (i=0,1,2,3)$ of SU(3) HB$\chi$PT. Thus,
$\bar{e}_i\,(i=14,15,16,17,18)$ are the five remaining dimension-four LECs that are relevant in our calculation. At fourth order, additional $\pi N$ amplitudes from one-loop diagrams must be taken into account. The pertinent one-loop diagrams generated by the interaction vertices from $\mathcal{L}_{\phi\phi}^{(2)}$, $\mathcal{L}_{\phi B}^{(1)}$ and $\mathcal{L}_{\phi B}^{(2)}$ are shown in Figure~\ref{fig:oneloopfeynman}. The expressions for these one-loop $\pi N$ amplitudes are rather tedious and will therefore not be reproduced here. The explicit analytical expressions for the loop contributions of order $\mathcal{O}(p^4)$ to elastic $\pi N$ scattering can be obtained from the authors upon request. For
orientation of the readers, we remark that these $\pi N$ loop amplitudes are composed of the following basic loop functions:
\begin{align}
\label{A1}
J_0(w,m)=&\frac{1}{i}\int\frac{d^{D}l}{(2\pi)^D}\frac{1}{(v\cdot l-w)(m^2-l^2)}
=\frac{w}{8\pi^2}\Big(1-2\text{ln}\frac{m}{\lambda}\Big)\nonumber\\
&+\begin{cases}
\dfrac{1}{4\pi^2}\sqrt{w^2-m^2}\text{ln}\dfrac{-w+\sqrt{w^2-m^2}}{m}& (w<-m),\\
-\dfrac{1}{4\pi^2}\sqrt{m^2-w^2}\text{arccos}\dfrac{-w}{m}& (-m<w<m),\\
\dfrac{1}{4\pi^2}\sqrt{w^2-m^2}\Bigg(i\pi-\text{ln}\dfrac{w+\sqrt{w^2-m^2}}{m}\Bigg)&(w>m),
\end{cases}
\end{align}
\begin{align}
\label{A2}
\frac{1}{i}\int\frac{d^{D}l}{(2\pi)^{D}}\frac{\{1,l^\mu,l^\mu l^\nu\}}{(m^2-l^2)[m^2-(l-k)^2]}=\{I_0(t,m),\frac{k^\mu}{2}I_0(t,m),g^{\mu\nu}I_2(t,m)+k^\mu k^\nu I_3(t,m)\},
\end{align}
\begin{align}
\label{A3}
I_0(t,m)=\frac{1}{8\pi^2}\Bigg\{\frac{1}{2}-\text{ln}\frac{m}{\lambda}-\sqrt{1-\frac{4m^2}{t}}\text{ln}\frac{\sqrt{4m^2-t}+\sqrt{-t}}{2m}\Bigg\},
\end{align}
\begin{align}
\label{A4}
I_2(t,m)=\frac{1}{48\pi^2}\Bigg\{2m^2-\frac{5t}{12}+\Big(\frac{t}{2}-3m^2\Big)\text{ln}\frac{m}{\lambda}-\frac{(4m^2-t)^{3/2}}{2\sqrt{-t}}\text{ln}\frac{\sqrt{4m^2-t}+\sqrt{-t}}{2m}\Bigg\},
\end{align}
\begin{align}
\label{A5}
I_3(t,m)=\frac{1}{24\pi^2}\Bigg\{\frac{7}{12}-\frac{m^2}{t}-\text{ln}\frac{m}{\lambda}-\Big(1-\frac{m^2}{t}\Big)\sqrt{1-\frac{4m^2}{t}}\text{ln}\frac{\sqrt{4m^2-t}+\sqrt{-t}}{2m}\Bigg\}.
\end{align}
Note that terms proportional to the divergent constant $\lambda^{
D-4}[\frac{1}{D-4}+{\frac{1}{2}}(\gamma_E-1-\ln 4\pi)]$ have been dropped. Furthermore, one observes that the one-loop amplitudes of order $\mathcal{O}(p^4)$ involve the full set of dimension-two LECs $b_i\,(i=D,F,0,...,11)$ and not just the few combinations of LECs $C_i\, (i=0,1,2,3)$ arising from tree diagrams.

\section{Partial-wave phase shifts and scattering lengths}
\label{phase}
The partial-wave amplitudes $f_{j}^{(I)}(\bm{q}^2)$, where $j=l\pm 1/2$ refers to the total angular momentum and $l$ to orbital angular momentum, are obtained
from the non-spin-flip and spin-flip amplitudes by the following projection formula:
\begin{align}
\label{eq32}
f_{l\pm 1/2}^{(I)}(\bm{q}^2)=\frac{M_{N}}{8\pi\sqrt{s}}\int_{-1}^{+1}dz\Big\{V_{\pi N}^{(I)}(w,t)\, P_{l}(z)+\bm{q}^{2}W_{\pi N}^{(I)}(w,t)[P_{l\pm 1}(z)-zP_{l}(z)]\Big\},
\end{align}
where $P_{l}(z)$ denotes the conventional Legendre polynomial, and $\sqrt{s}=\sqrt{m_\pi^2+\bm{q}^2}+\sqrt{M_N^2+\bm{q}^2}$ is the total center-of-mass energy. For the energy range considered in this paper, the phase shifts $\delta_{l\pm 1/2}^{(I)}$ are calculated as (see also refs.~\cite{gass1991,fett1998})
\begin{align}
\label{eq33}
\delta_{l\pm 1/2}^{(I)}(\bm{q}^2)=\text{arctan}\Big[|\bm{q}|\,\text{Re}\,f_{l\pm 1/2}^{(I)}(\bm{q}^2)\Big].
\end{align}

The scattering lengths for $S$-waves and the scattering volumes for $P$-waves are obtained by dividing out the threshold behavior of the respective partial-wave amplitude and approaching the threshold \cite{eric1988}
\begin{align}
\label{eq34}
a_{l\pm 1/2}^{(I)}=\lim\limits_{|\bm{q}| \rightarrow 0}|\bm{q}|^{-2l}f_{l\pm 1/2}^{(I)}(\bm{q}^2).
\end{align}

\section{Results and discussion}
\label{results}
Before presenting results, we have to determine the LECs. The parameters $M_0$, $b_D$, $b_F$ and $b_0$ have been determined in our previous paper by using of the four octet-baryon masses ($M_N$, $M_\Sigma$, $M_\Xi$, $M_\Lambda$) and the pion-nucleon $\sigma$ term $\sigma_{\pi N}=59.1\pm 3.5$ MeV\cite{huan2015}. In this paper, we take the same value $M_0=646.3$ MeV, but the parameters $b_0$, $b_D$, and $b_F$ are obtained in the fitting. Unfortunately, the 14 dimension-two LECs cannot be regrouped up to fourth order in SU(3) HB$\chi$PT. Then, in total, there are 23 unknown constants that need to be determined. Throughout this paper, we also use $m_\pi=139.57 \,\text{MeV}$, $m_K=493.68 \,\text{MeV}$, $m_\eta=547.86 \, \text{MeV}$, $f_\pi=92.07 \, \text{MeV}$, $f_K=110.03 \, \text{MeV}$, $f_\eta=1.2 f_\pi$, $M_N=938.92\pm 1.29 \, \text{MeV}$, $M_\Sigma=1191.01\pm 4.86\,\text{MeV}$, $M_\Xi=1318.26\pm 6.30\,\text{MeV}$, $M_\Lambda=1115.68\pm 5.58\,\text{MeV}$, $\lambda=4\pi f_\pi=1.16\,\text{GeV}$, $D=0.80$, and $F=0.47$ \cite{pdg2018,chan2018,mark2019}.

\begin{table*}[!b]
\centering
\begin{threeparttable}
\caption{\label{fittingresult}Values of the various fits. The fits N3LO, N2LO, and NLO refer to the best fit up to fourth, third, and second order, respectively. For a detailed description of these fits, see the main text. Note that the $(*)$ values are calculated by $b_{i}$.}
\begin{tabular}{ccccccc}
\midrule
\toprule
 & N3LO & N2LO & NLO &\\
\midrule
$b_D$ ($\text{GeV}^{-1}$)&$-3.44\pm 0.13$&&&\\
\midrule
$b_F$ ($\text{GeV}^{-1}$)&$4.64\pm 0.18$&&&\\
\midrule
$b_0$ ($\text{GeV}^{-1}$)&$-1.03\pm 0.25$&&&\\
\midrule
$b_1$ ($\text{GeV}^{-1}$)&$1.61\pm 0.08$&&&\\
\midrule
$b_2$ ($\text{GeV}^{-1}$)&$0.10\pm 0.03$&&&\\
\midrule
$b_3$ ($\text{GeV}^{-1}$)&$-4.50\pm 0.32$&&&\\
\midrule
$b_4$ ($\text{GeV}^{-1}$)&$-1.34\pm 0.17$&&&\\
\midrule
$b_5$ ($\text{GeV}^{-1}$)&$-23.28\pm 1.56$&&&\\
\midrule
$b_6$ ($\text{GeV}^{-1}$)&$1.90\pm 0.21$&&&\\
\midrule
$b_7$ ($\text{GeV}^{-1}$)&$8.96\pm 0.99$&&&\\
\midrule
$b_8$ ($\text{GeV}^{-1}$)&$-5.62\pm 0.67$&&&\\
\midrule
$b_9$ ($\text{GeV}^{-1}$)&$1.86\pm 0.23$&&&\\
\midrule
$b_{10}$ ($\text{GeV}^{-1}$)&$-0.38\pm 0.00$&&&\\
\midrule
$b_{11}$ ($\text{GeV}^{-1}$)&$17.66\pm 1.56$&&&\\
\midrule
$C_0$ ($\text{GeV}^{-1}$)&$-0.86\pm 0.55^{(*)}$&$-3.59\pm 0.06$&$-1.69\pm 1.49$&\\
\midrule
$C_1$ ($\text{GeV}^{-1}$)&$-7.31\pm 0.65^{(*)}$&$-6.60\pm 0.03$&$-4.30\pm 0.15$&\\
\midrule
$C_2$ ($\text{GeV}^{-1}$)&$-3.46\pm 2.53^{(*)}$&$3.71\pm 0.05$&$2.94\pm 1.45$&\\
\midrule
$C_3$ ($\text{GeV}^{-1}$)&$1.48\pm 0.23^{(*)}$&$1.41\pm 0.01$&$1.37\pm 0.07$&\\
\midrule
$H_1$ ($\text{GeV}^{-2}$)&$40.38\pm 4.12$&$6.26\pm 0.35$&&\\
\midrule
$H_2$ ($\text{GeV}^{-2}$)&$10.98\pm 1.52$&$4.78\pm 0.07$&&\\
\midrule
$H_3$ ($\text{GeV}^{-2}$)&$-9.88\pm 0.82$&$-7.65\pm 0.31$&&\\
\midrule
$H_4$ ($\text{GeV}^{-2}$)&$-15.48\pm 0.78$&$-5.76\pm 0.15$&&\\
\midrule
$\bar{e}_{14}$ ($\text{GeV}^{-3}$)&$0.99\pm 0.03$&&&\\
\midrule
$\bar{e}_{15}$ ($\text{GeV}^{-3}$)&$-26.22\pm 3.12$&&&\\
\midrule
$\bar{e}_{16}$ ($\text{GeV}^{-3}$)&$22.61\pm 2.93$&&&\\
\midrule
$\bar{e}_{17}$ ($\text{GeV}^{-3}$)&$4.57\pm 0.52$&&&\\
\midrule
$\bar{e}_{18}$ ($\text{GeV}^{-3}$)&$4.57\pm 0.52$&&&\\
\midrule
$\chi^2/\text{d.o.f.}$&$0.24$&$1.24$&$21.83$&\\
\bottomrule
\midrule
\end{tabular}
\end{threeparttable}
\end{table*}

 \begin{figure}[!t]
\centering
\includegraphics[height=7.5cm,width=12.75cm]{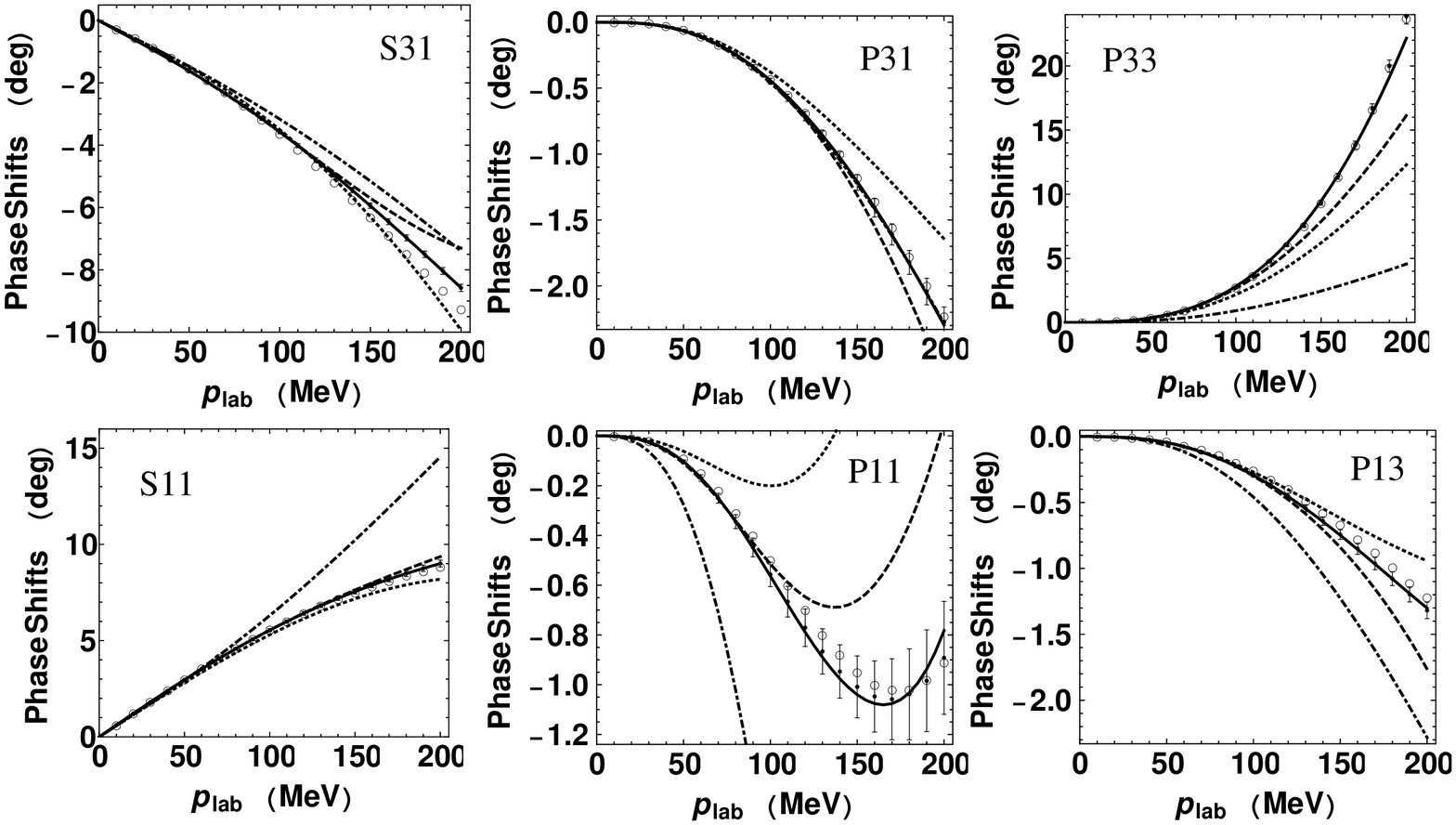}
\caption{\label{fig:pinphaseshifts}Fits and predictions for pion-nucleon ($\pi N$) phase shifts from RS equations versus the pion lab. momentum $|\bm{p}_{\text{lab}}|$ in $\pi N$ scattering. The dot-dashed, dotted, dashed, and solid lines refer to the best fits up to first, second, third, and fourth order, respectively. The open circles denote the WI08 solution, and the data with error bars are from RS equations. }
\end{figure}

\begin{table*}[!b]
\centering
\begin{threeparttable}
\caption{\label{thresholdparameterone}
Values of the $S$- and $P$-wave scattering lengths (fm) and scattering volumes ($\text{fm}^3$) in comparison with the values of the various analyses.}
\begin{tabular}{cccccccccccc}
\midrule
\toprule
 & Our results & SU(2) & SP98 & RS & EXP2001  & EXPnew\\
\midrule
$a_{0+}^{3/2}$ & $-0.122(4)$ & $-0.119$ & $-0.125(2)$ & $-0.122(3)$ & $-0.125(3) $& $-0.119(6)$ &\\
\midrule
$a_{0+}^{1/2}$ & $0.240(3)$ & $0.249$ & $0.250(2) $ & $0.240(2)$ & $0.250^{+0.006}_{-0.004}$& $0.235(2)$ &\\
\midrule
 $a_{1+}^{3/2}$ & $0.603(13)$ & $0.586$ & $0.595(5)$ & $0.598(8)$ & ... & ...  &\\
\midrule
 $a_{1+}^{1/2}$ & $-0.102(6)$ & $-0.054$ & $-0.038(8)$ &$-0.083(3)$ & ... & ...  &\\
  \midrule
 $a_{1-}^{3/2}$ & $-0.110(9)$ & $-0.113$ & $-0.122(6)$ &$-0.116(3)$ & ... & ...  &\\
\midrule
 $a_{1-}^{1/2}$ & $-0.221(12)$ & $-0.181$ & $-0.207(7)$ &$-0.200(11)$ & ... & ...  &\\
\bottomrule
\midrule
\end{tabular}
\end{threeparttable}
\end{table*}

 We have various fitting strategies to determine the pertinent constants, and the related discussion is given in ref.~\cite{huan2020}. One of them is using the octet-baryon masses ($M_{N,\Sigma,\Xi,\Lambda}$) and the phase shifts of $\pi N$ scattering simultaneously. This can no longer be done up to fourth order due to the appearance of the more constants for octet-baryon masses. Thus, we determine the LECs by using the $\pi N$ phase shifts from the RS equations \cite{hofe20152,hofe2016}. The explicit numerical solutions for the $\pi N$ phase shifts can be found in Appendix D of ref.~\cite{hofe2016}. Note that, the RS phase shifts are different from the WI08 phase shifts \cite{SAID,work2012} and the phase shifts errors can be obtained in RS solution. Thus, we do not need to choose an inaccurate uncertainty for all phase shifts before the fitting procedure. The choice for the RS phase shifts is also expected to improve our fitting. The data points of the $S$ and $P$ waves in the range of 10-200 MeV pion lab. momentum are used. The resulting LECs can be found in the N3LO of Table~\ref{fittingresult}. The uncertainty for the respective parameter is purely estimated (for a detailed discussion, see, e.g., refs.\cite{doba2014,carl2016}). It is not surprising that most LECs are natural size $\mathcal{O}(1)$; i.e., the absolute values of these LECs are between one and ten when one introduces the dimensionless LECs (e.g., $b^{'}_i=2M_0b_i$), whereas some of LECs come out fairly large. The situation also can be found in SU(2) HB$\chi$PT \cite{fett2000}. In fact, the dimension three LECs $H_i$ are not terribly large. Because we have $H_1=4\bar{d}_5+2(\bar{d}_1+\bar{d}_2)$, $H_2=\bar{d}_1+\bar{d}_2$, $H_3=2\bar{d}_3$, $H_4=\bar{d}_{14}-\bar{d}_{15}$, where $\bar{d}_i$ are from SU(2) HB$\chi$PT. Then, we can obtain $\{\bar{d}_1+\bar{d}_2, \bar{d}_3, \bar{d}_5, \bar{d}_{14}-\bar{d}_{15}\}=\{10.98, -4.94, 4.61, 15.48\} \text{GeV}^{-2} $. However, the results of $\bar{d}_i$ are clearly not $\mathcal{O}(1)$, which indicates the presence of additional degree of freedom. For instance, the presence of the $\Delta(1232)$ resonance can explain the values of some LECs \cite{bern19951,bern1997,bech1999,hofe20151}. The corresponding $S$- and $P$-wave phase shifts are shown by the solid lines of Fig.~\ref{fig:pinphaseshifts}. Obviously, we obtain an excellent description of all waves. Especially, the description of the $P11$-, $P33$-, and $P13$-wave phase shifts are improved at high energies as compared to the third-order calculation. For comparison, we present results from the best fits up to third (N2LO) and second (NLO) order in Table~\ref{fittingresult}. Clearly, the resulting LECs have different values in the respective order. It is not surprising that the LECs at NLO and N2LO show different values. However, the LEC values at N3LO and N2LO are also very different, even an order of magnitude in some cases, which are expected to be same. Thus, for the amplitudes up to the given order, the corresponding fit to determine their respective constants is necessary. This is similar to what was found in SU(2) HB$\chi$PT \cite{hofe20151,siem2017}. The feature can lead to the potential convergence problems in the chiral expansion; further analysis of the threshold and subthreshold parameters is required.

\begin{table*}[!t]
\centering
\begin{threeparttable}
\caption{\label{thresholdparametertwo}
Values of the $S$- and $P$-wave scattering lengths (fm) and scattering volumes ($\text{fm}^3$) from the different order.}
\begin{tabular}{cccccccccccc}
\midrule
\toprule
 & $\mathcal{O}(q)$ &  $\mathcal{O}(q^2)$ &  $\mathcal{O}(q^3)$ & $\mathcal{O}(q^4)$& \\
\midrule
$a_{0+}^{3/2}$  & $-0.113$ & $-0.111(87)$ & $-0.121(18)$ & $-0.122(4)$ &\\
\midrule
$a_{0+}^{1/2}$  & $0.225$ & $0.226(87)$ & $0.240(16) $ & $0.240(3)$ &\\
\midrule
 $a_{1+}^{3/2}$  & $0.241$ & $0.531(11)$ & $0.624(15)$ & $0.603(13)$ &\\
\midrule
 $a_{1+}^{1/2}$  & $-0.121$ & $-0.082(9)$ & $-0.080(11)$ & $-0.102(6)$& \\
  \midrule
 $a_{1-}^{3/2}$  & $-0.121$ & $-0.106(11)$ & $-0.113(14)$ & $-0.110(9)$ & \\
\midrule
 $a_{1-}^{1/2}$  & $-0.483$ & $-0.129(15)$ & $-0.215(19)$ & $-0.221(12)$ & \\
\bottomrule
\midrule
\end{tabular}
\end{threeparttable}
\end{table*}

\begin{table*}[!b]
\centering
\begin{threeparttable}
\caption{\label{subthreshold}Subthreshold parameters for the $D^{\pm}$ and $B^{\pm}$ amplitudes in comparison to the RS analysis.}
\begin{tabular}{ccccccc}
\midrule
\toprule
                         & N2LO & N3LO & RS &\\
\midrule
$d_{00}^{+}[m_\pi^{-1}]$ &$-0.91(12)$   &$-0.99(15)$  &  $-1.36(3)$&\\
\midrule
$d_{10}^{+}[m_\pi^{-3}]$ &$1.12(7)$     &$1.21(13)$   &  $1.16(2)$ &\\
\midrule
$d_{01}^{+}[m_\pi^{-3}]$ &$1.30(5)$     &$1.28(8)$    &   $1.16(2)$ &\\
\midrule
$d_{20}^{+}[m_\pi^{-5}]$ &$-0.13(6)$    &$0.13(6)$   &  $0.196(3)$ &\\
\midrule
$d_{11}^{+}[m_\pi^{-5}]$ &$-0.011(9)$   &$0.16(11)$   &  $0.185(3)$ &\\
\midrule
$d_{02}^{+}[m_\pi^{-5}]$ &$0.0298(18)$  &$0.076(13)$  &  $0.0336(6)$ &\\
\midrule
$b_{00}^{+}[m_\pi^{-3}]$ &$-6.14(8)$    &$-6.11(14)$  &  $-3.45(7)$ &\\
\midrule
$d_{00}^{-}[m_\pi^{-2}]$ &$0.93(11)$    &$0.63(17)$   &  $1.41(1)$ &\\
\midrule
$d_{10}^{-}[m_\pi^{-4}]$ &$-0.395(21)$   &$-0.12(9)$  &  $-0.159(4)$ &\\
\midrule
$d_{01}^{-}[m_\pi^{-4}]$ &$-0.105(13)$   &$-0.24(9)$  &  $-0.141(5)$ &\\
\midrule
$b_{00}^{-}[m_\pi^{-2}]$ &$9.92(6)$      &$9.73(8)$   &  $10.49(11)$ &\\
\midrule
$b_{10}^{-}[m_\pi^{-4}]$ &$0.44(5)$      &$0.11(14)$  &  $1.00(3)$ &\\
\midrule
$b_{01}^{-}[m_\pi^{-4}]$ &$-0.28(9)$     &$0.39(16)$  &  $0.21(2)$ &\\
\bottomrule
\midrule
\end{tabular}
\end{threeparttable}
\end{table*}

\setlength{\tabcolsep}{1.7mm}{
\begin{table*}[!t]
\centering
\begin{threeparttable}
\caption{\label{LECsSU3vsSU2}Results for the $\pi N$ LECs at NLO and N2LO in both SU(3) and SU(2) HB$\chi$PT. The results for the $c_i$ and $\bar{d}_i$ are given in units of $\text{GeV}^{-1}$ and $\text{GeV}^{-2}$, respectively. Note that, the results from SU(2) and SU(3) are the same at NLO.}
\begin{tabular}{ccccccccccccccccc}
\midrule
\toprule
&&&&NLO&&&&&\\
\midrule
  & $c_1$     & $c_2$     & $c_3$      & $c_4$   &  & &  &  &\\
\midrule
SU(2) &$-0.75(4)$ &$1.81(3)$  &  $-3.62(6)$ & $2.16(3)$ && & & &\\
\midrule
&&& & N2LO & & & & &\\
\midrule
      & $c_1$     & $c_2$     & $c_3$      & $c_4$   & $\bar{d}_1$+$\bar{d}_2$  & $\bar{d}_3$  & $\bar{d}_5$  & $\bar{d}_{14}$-$\bar{d}_{15}$ &\\
\midrule
SU(3) &$-1.23(4)$ &$4.57(3)$  &  $-6.36(6)$ & $4.19(3)$ &$1.08(6)$ &$-0.28(2)$ &$0.04(4)$ &$-1.56(6)$ &\\
\midrule
$\text{SU(3)}_K$ &$-1.23(4)$ &$4.54(3)$  &  $-6.34(6)$ & $4.16(3)$ &$1.11(6)$ &$-0.29(2)$ &$0.04(4)$ &$-1.57(6)$ &\\
\midrule
SU(2) &$-1.07(4)$ &$3.20(3)$  &  $-5.32(6)$ & $3.55(3)$ &$1.03(6)$ &$-0.47(2)$ &0.13(4)   &$-1.89(6)$ &\\
\bottomrule
\midrule
\end{tabular}
\end{threeparttable}
\end{table*}}

 In the following, let us apply the chiral fourth-order amplitudes to estimate the threshold parameters. The analytical expressions for scattering lengths are shown in Appendix~\ref{scattering lengths}. However, the pion-nucleon scattering lengths and scattering volumes can also be obtained by using an incident pion lab. momentum $|\bm{p}_{\text{lab}}|=5\,\text{MeV}$ and approximating their values at the threshold. The results are shown in Table~\ref{thresholdparameterone} in comparison with the values of the various analyses. Obviously, our results for most of the threshold parameters are consistent with the ones from SU(2) HB$\chi$PT and SP98 in ref.~\cite{fett2000}, and the RS solution in ref.~\cite{hofe2016}. For scattering volume $a_{1+}^{1/2}$, our result is larger than the SU(2) and SP98, but it is closer to the RS solution. The values of SP98 are obtained by the use of dispersion relations with the help of a fairly precise tree-level model. In addition, there are two experimental values for scattering lengths in Table~\ref{thresholdparameterone}. The values of EXP2001 are from pionic hydrogen and deuterium in ref.~\cite{schr2001}. The values are updated in ref.~\cite{hofe2016} and are not shown again because they are very close to the results from RS solution. The latter, EXPnew, are obtained by combining with the analysis of the results from refs.~\cite{baru2011,baru20112,hofe2012,henn2014,hofe20152,elvi2017}. As expected, our results for scattering lengths are consistent with those values within errors. In order to check the convergence, the threshold parameters from the different order are shown in Table~\ref{thresholdparametertwo}.

 Now we discuss the subthreshold parameters. For comparing with the precise subthreshold parameters from the RS analysis \cite{hofe20151}, we consider the parameters from $D^{\pm}$ and $B^{\pm}$ amplitudes. Here, $D^{\pm}=A^{\pm}+\nu B^{\pm}$, for the detailed relations about them, see Eqs.~(3.32)-(3.34) of ref.~\cite{hofe2016}. We can obtain the subthreshold parameters from our $V^{I}(w,t)$ and $W^{I}(w,t)$ amplitudes by following Appendix E of ref.~\cite{fett1998}. Note that, we have $\{V,W\}^{3/2}=\{g,h\}^{+}-\{g,h\}^{-}$ and $\{V,W\}^{1/2}=\{g,h\}^{+}+2\{g,h\}^{-}$ for the isospin relationship. The corresponding subthreshold parameters can be obtained by expanding our amplitudes around ($v=0, t=0$). We show the results from our amplitudes with the LECs extracted from the phase shifts up to third order (N2LO) and fourth order (N3LO) in comparison to the RS analysis in Table~\ref{subthreshold}. Not surprisingly, the values from the third-order amplitudes are not precise. When using the fourth order amplitudes, we do not find an improvement of the subthreshold parameters, and in some parameters they are even clearly worse. The situation is same as in SU(2) HB$\chi$PT \cite{fett1998,fett2000}. Furthermore, the convergence problems obviously come out. Most of the subthreshold parameters from N3LO and N2LO are not consistent, and some even have opposite signs. The fact is that one cannot properly reproduce the subthreshold kinematic parameters from the amplitudes with convergence problems in the low-energy region. This is exactly similar to what is found when the scattering lengths are predicted from LECs extracted at the subthreshold point up to N2LO. Thus, we can use the subthreshold parameters to determine the LECs, as done in ref.~\cite{hofe20151}, and then discuss the pertinent physical quantity. But we have 23 LECs at fourth order in SU(3) HB$\chi$PT. That means we need 23 precise subthreshold parameters from RS analysis. This solution will be done in future works.

 However, the subthreshold parameters only involve eight LECs at third order in both SU(3) and SU(2) HB$\chi$PT. That means we need the eight subthreshold parameters to determine the LECs at third order. The expressions of the subthreshold parameters in SU(3) HB$\chi$PT can be found in Appendix~\ref{subthresholdexpressions}. The pertinent expressions in SU(2) HB$\chi$PT can be obtained by subtracting the contributions from $K$ and $\eta$ internal meson lines in one-loop diagrams. We use the precise values of the subthreshold parameters from the RS analysis \cite{hofe20151} to determine the same LECs in both SU(3) and SU(2) HB$\chi$PT. The results are shown in Table~\ref{LECsSU3vsSU2}. The results of SU(2) are slightly different from ref.~\cite{hofe20151} because of the different physical quantity from PDG was taken. Note that, the errors as propagated from the subthreshold parameters are same in both SU(3) and SU(2) solutions because the expressions from $K$ and $\eta$ loop diagrams do not involve the LECs at third order. The $\text{SU(3)}_K$ presents the result that the contributions from $\eta$-meson loop diagrams is subtracted. Clearly, all LECs, except for $\bar{d}_1+\bar{d}_2$, are obviously different in SU(2) and SU(3) solutions. Thus, it is interesting to predict the threshold parameters by using the LECs from the subthreshold parameters. Let us look at the scattering lengths as an example; for the pertinent expressions, see Appendix~\ref{scattering lengths}. In unites of fm, we obtain
 \begin{align}
\label{eq16}
&a_{0+}^{3/2}=\{-0.113,-0.146,-0.130\}_{\text{SU(2)}},\,\{-0.113,-0.146,-0.128\}_{\text{SU(3)}},\nonumber\\
&a_{0+}^{1/2}=\{0.225,0.192,0.263\}_{\text{SU(2)}},\,\{0.225,0.192,0.266\}_{\text{SU(3)}},
\end{align}
where the three entries of the arrays refer to the LO, NLO, and N2LO scattering length values, respectively. The situation $\text{SU(3)}_K$ is not shown because the contributions from $\eta$-meson loop are very small. Not surprisingly, the results from the SU(2) and SU(3) are almos the same at third order. The contributions from the kaon loops cannot improve the predictions for the scattering lengths from LECs extracted at the subthreshold point up to N2LO. Therefore, the convergence problems still exist the same as in SU(2) HB$\chi$PT \cite{hofe20151}. The differences between the scattering lengths from eq.~(\ref{eq16}) and Table~\ref{thresholdparametertwo} emphasize these convergence problems. Furthermore, we try to impose constraints to the LECs both from the threshold and subthreshold kinematic parameters to analyze whether the discrepancies improve. First, we obtain the LECs combinations $-2c_1+c_2+c_3=0.59 \,\text{GeV}^{-1}$ and $\bar{d}_1+\bar{d}_2+\bar{d}_3+2\bar{d}_5=0.28\,\text{GeV}^{-2}$ by using the two scattering lengths from the RS solution at N2LO in SU(3) HB$\chi$PT. Then, the $c_2$, $c_3$, $\bar{d}_1+\bar{d}_2$, and $\bar{d}_3$ are fixed by the subthreshold parameters; see Table~\ref{LECsSU3vsSU2}. Now we predict the subthreshold parameters $d_{00}^{+}=-1.41\,[m_\pi^{-1}]$ and $d_{00}^{-}=1.30\,[m_\pi^{-2}]$. The $d_{00}^{+}$ is very closed to the value from the RS solution, but the $d_{00}^{-}$ has a certain gap. The results are disappointing. However, the inclusion of the resonance $\Delta(1232)$ might help to solve all previous observed problems.

 Finally, we discuss the convergence of the chiral expansion in detail. In Fig.~\ref{fig:pinphaseshifts}, we show the best fits up to the respective order. In all partial waves, the fourth-order corrections are smaller than the third-order ones below 150 MeV, indicating convergence. In most cases, the third-order corrections are smaller than the second-order ones. The second-order corrections are smaller than the first-order ones in a few partial waves. Therefore, the convergence of the chiral expansion becomes better along with the increase of the order. However, with the energy increasing the convergence of the chiral expansion becomes worse. It is not surprising because the chiral expansion is expanded in terms of $p/\Lambda_{\chi}$. We can also study the convergence of the threshold parameters; see Table~\ref{thresholdparametertwo}. The two scattering lengths $a_{0+}^{3/2}$ and $a_{0+}^{1/2}$ from $S$ waves are almost unchanged in a different order, indicating convergence in each order. The other four scattering volumes from $P$ waves are consistent from order $\mathcal{O}(q^3)$. The calculation results indicate that the convergence of the threshold parameters is fast because it is due to $m_\pi/\Lambda_{\chi}\sim 1/7$. However, the convergence problems obviously come out when the subthreshold parameters are considered. These problems become more complicated. A great improvement on the convergence of the chiral expansion is still necessary.

 In summary, we calculated the $T$ matrices for pion-nucleon scattering to the fourth order in SU(3) HB$\chi$PT. We fitted the RS phase shifts of $\pi N$ scattering in range of $10-200$ MeV pion lab. momentum to determine the LECs. This led to an good description of $S$- and $P$-wave phase shifts below 200 MeV pion lab. momentum. The separated LECs $b_i$ were also obtained, they can be used as input to the other physical processes. The scattering lengths and scattering volumes were also calculated at this order, which turned out to be in agreement with those of other approaches and available experimental data. Then, we discussed the subthreshold parameters, and found that it was not a good choice to obtain the subthreshold parameters from the $\pi N$ phase shifts. We also discussed the different LECs obtained by subthreshold parameters in both SU(2) and SU(3) HB$\chi$PT, and it turned out that the contribution from kaon loops can contribute to the LECs, but the prediction from subthreshold parameters to threshold parameters was not improved. Finally, the convergence of the chiral expansion was discussed in detail. To sum up, the calculation of $\pi N$ scattering in SU(3) HB$\chi$PT provides the possibility to consider explicitly more complex processes involving strangeness. An improved result for $\pi N$ scattering can be achieved through including the resonance $\Delta(1232)$ and the other hadronic contributions.

\section*{Acknowledgments}
This work is supported by the National Natural Science Foundation of China under Grant No. 11947036. I thank Norbert Kaiser (Technische Universit\"{a}t M\"{u}nchen), Yan-Rui Liu (Shandong University), Li-Sheng Geng (Beihang University), Jun-Xu Lu (Beihang University), and Jing Ou-Yang (Jishou University) for very helpful discussions.

\appendix\markboth{Appendix}{Appendix}
\renewcommand{\thesection}{\Alph{section}}
\numberwithin{equation}{section}
\section{Scattering lengths}
\label{scattering lengths}
In this appendix, the analytical expressions for the scattering lengths from SU(3) HB$\chi$PT up to third order are given by
\begin{align}
\label{A1}
a_{0+}^{+}=&\frac{m_\pi^2[-(D+F)^2+8M_N(-2c_1+c_2+c_3)+4m_\pi^2(D+F)\bar{d}_{18}]}{16\pi(M_N+m_\pi)f_\pi^2}+\frac{3(D+F)^2M_N m_\pi^3}{256\pi^2(M_N+m_\pi)f_\pi^4}\nonumber\\
&-\frac{M_N m_\pi^2\sqrt{m_K^2-m_\pi^2}}{384\pi^3(M_N+m_\pi)f_\pi^2f_K^2}(7\pi+4\text{arccos}
\frac{m_\pi}{m_K}+4\text{arcsin}\frac{m_\pi}{m_K})-\frac{(D-3F)^2 M_N m_\pi^2 m_\eta}{768\pi^2(M_N+m_\pi)f_\pi^2f_\eta^2},
\end{align}
\begin{align}
\label{A1}
a_{0+}^{-}=&\frac{M_N m_\pi}{8\pi (M_N+m_\pi)f_\pi^2}+\frac{m_\pi^3[(D+F)^2+32M_N^2(\bar{d}_1+\bar{d}_2+\bar{d}_3+2\bar{d}_5)]}{32\pi M_N(M_N+m_\pi)f_\pi^2}+\frac{M_N m_\pi^3}{64\pi^3(M_N+m_\pi)f_\pi^4}\nonumber\\
&-\frac{M_N m_\pi^2}{384\pi^3 (M_N+m_\pi)f_\pi^2 f_K^2}[-3m_\pi+\sqrt{m_K^2-m_\pi^2}(\pi-2\text{arccos}\frac{m_\pi}{m_K}+4\text{arcsin}\frac{m_\pi}{m_K})],
\end{align}
where we use the physical decay constants $f_{\pi}^2f_{K}^2$ and $f_{\pi}^2f_{\eta}^2$ in $K$ and $\eta$ internal meson lines from one-loop diagrams, respectively. Note that, for comparing with the results from SU(2) HB$\chi$PT, the same LECs ($c_i,\,\bar{d}_i$) are used. For the isospin relationship, we also have $a_{0+}^{3/2}=a_{0+}^{+}-a_{0+}^{-}$ and $a_{0+}^{1/2}=a_{0+}^{+}+2a_{0+}^{-}$.
\section{Subthreshold parameters}
\label{subthresholdexpressions}
In this appendix we present the expressions for the subthreshold parameters from the third order amplitudes in SU(3) HB$\chi$PT. In the same conventions with the scattering lengths, the subthreshold parameters read
\begin{align}
\label{A1}
d_{00}^{+}=&-\frac{2(2c_1-c_3)m_\pi^2}{f_\pi^2}+\frac{(D+F)^2[3+8(D+F)^2]m_\pi^3}{64\pi f_\pi^4}+\frac{1}{192\pi f_\pi^2f_K^2}(19D^4+12D^3F\nonumber\\
&+58D^2F^2-36DF^3+75F^4)m_K m_\pi^2+\frac{1}{192\pi f_\pi^2f_\eta^2}(D-3F)^2[-1+4(D+F)^2]m_\eta m_\pi^2,
\end{align}
\begin{align}
\label{A2}
d_{10}^{+}=&\frac{2c_2}{f_\pi^2}-\frac{[4+5(D+F)^4]m_\pi}{32\pi f_\pi^4}-\frac{1}{768\pi f_\pi^2f_K^2 m_K}[72m_K^2+(19D^4+12D^3F+58D^2F^2\nonumber\\
&-36DF^3+75F^4)(m_\pi^2+4m_K^2)]-\frac{1}{192\pi f_\pi^2 f_\eta^2 m_\eta}(D-3F)^2(D+F)^2(m_\pi^2+4m_\eta^2),
\end{align}
\begin{align}
\label{A3}
d_{01}^{+}=&-\frac{c_3}{f_\pi^2}-\frac{(D+F)^2[77+48(D+F)^2]m_\pi}{768\pi f_\pi^4}-\frac{1}{384\pi f_\pi^2f_K^2}(15D^2-18DF+27F^2+19D^4+75F^4\nonumber\\
&+58D^2F^2-36DF^3+12D^3F)m_K+\frac{1}{6912\pi f_\pi^2 f_\eta^2 m_\eta}(D-3F)^2[5m_\pi^2-72(D+F)m_\eta^2],
\end{align}
\begin{align}
\label{A4}
d_{00}^{-}=&\frac{1}{2f_\pi^2}+\frac{4m_\pi^2(\bar{d}_1+\bar{d}_2+2\bar{d}_5)}{f_\pi^2}+\frac{(D+F)^4m_\pi^2}{48\pi^2f_\pi^4}
+\frac{1}{288\pi^2f_\pi^2f_K^2}(23D^4-20D^3F-14D^2F^2\nonumber\\
&+60DF^3+111F^4)m_\pi^2+\frac{1}{36\pi^2f_\pi^2f_\eta^2}(D-3F)^2(D+F)^2m_\pi^2,
\end{align}
\begin{align}
\label{A5}
d_{10}^{-}=&\frac{4\bar{d}_3}{f_\pi^2}-\frac{15+7(D+F)^4}{240\pi^2f_\pi^4}-\frac{1}{1440\pi^2f_\pi^2f_K^2m_K^2}[5(9+23D^4-20D^3F-14D^2F^2
+60DF^3\nonumber\\
&+111F^4)m_K^2+2(17D^4-12D^3F-2D^2F^2+36DF^3+57F^4)m_\pi^2]\nonumber\\
&-\frac{1}{180\pi^2f_\pi^2f_\eta^2m_\eta^2}(D-3F)^2(D+F)^2(m_\pi^2+5m_\eta^2),
\end{align}
\begin{align}
\label{A5}
d_{01}^{-}=&-\frac{2(\bar{d}_1+\bar{d}_2)}{f_\pi^2}-\frac{1+7(D+F)^2+2(D+F)^4}{192\pi^2f_\pi^4}+\frac{1}{3456\pi^2f_\pi^2f_K^2}[39-47D^2+282DF
+141F^2\nonumber\\
&-138D^4+120D^3F+84D^2F^2-360DF^3-666F^4]-\frac{1}{72\pi^2f_\pi^2f_\eta^2}(D-3F)^2(D+F)^2,
\end{align}
\begin{align}
\label{A5}
b_{00}^{+}=&\frac{4M_N(\bar{d}_{14}-\bar{d}_{15})}{f_\pi^2}-\frac{(D+F)^4M_N}{8\pi^2f_\pi^4}-\frac{1}{432\pi^2f_\pi^2f_K^2}(47D^4+60D^3F+18D^2F^2
+108DF^3\nonumber\\
&+279F^4)M_N-\frac{1}{24\pi^2f_\pi^2f_\eta^2}(D-3F)^2(D+F)^2M_N,
\end{align}
\begin{align}
\label{A5}
b_{00}^{-}=&\frac{1}{2f_\pi^2}+\frac{2c_4 M_N}{f_\pi^2}-\frac{(D+F)^2[1+(D+F)^2]M_N m_\pi}{8\pi f_\pi^4}-\frac{1}{96\pi f_\pi^2f_K^2}(2D^2-12DF-6F^2+9D^4\nonumber\\
&+20D^3F-2D^2F^2+36DF^3+33F^4)M_N m_K-\frac{1}{48\pi f_\pi^2 f_\eta^2}(D-3F)^2(D+F)^2M_N m_\eta.
\end{align}

\bibliographystyle{unsrt}
\bibliography{latextemplate}

\end{document}